# Physico-thermal and geochemical behavior and alteration of the Au indicator gangue hydrothermal quartz at the Kubi Gold Ore Deposits

Gabriel K. Nzulu, Lina Rogström, Jun Lu, Hans Högberg, Per Eklund, Lars Hultman, and Martin Magnuson

*Department of Physics, Chemistry and Biology (IFM) Linköping University, Sweden*

**Abstract**

Altered and gangue quartz in hydrothermal veins from the Kubi Gold deposit in Dunkwa on Offin in the central region of Ghana are investigated for possible Au-associated indicator minerals and to provide the understanding and increase the knowledge of the mineral hosting and alteration processes in quartz. X-ray diffraction, air annealing furnace, differential scanning calorimetry, energy dispersive X-ray spectroscopy, and transmission electron microscopy have been applied on different quartz types outcropping from surface and bedrocks at the Kubi Gold Mining to reveal the material properties at different temperatures. From the diffraction results of the fresh and annealed quartz samples, we find that the samples contain indicator and the impurity minerals iron disulfide, biotite, titanium oxide, and magnetite. These minerals, under oxidation process between 574-1400 °C temperatures experienced hematite alterations and a transformation from α-quartz to β-quartz and further to cristobalite as observed from the calorimetry scans for hydrothermally exposed materials. The energy dispersive spectroscopy revealed elemental components of Fe, S, Mg, K, Al, Ti, Na, Si, O, and Ca contained in the samples, and these are attributed to the impurity phase minerals observed in the diffraction. The findings also suggest that during the hydrothermal flow regime, impurity minerals and metals can be trapped by voids and faults. Under favorable temperature conditions, the trapped minerals can be altered to change color at different depositional stages by oxidation and reduction processes leading to hematite alteration which is a useful indicator mineral in mineral exploration.

1. Introduction

Quartz, silicon oxide mineral is considered the second most abundant gangue mineral on earth, formed from the Earth's crust and down to the core of the earth and crystallizes or solidifies depending on conditions such as pressure, stress, temperature, temperature gradients, time, presence, and composition of magma (melts and fluids). It also depends on geological activities *i.e.*, deformations (cracks and voids) at different stages of hydrothermal systems (temperature between 350-800 °C and lower) [1, 2], and is very abundant in the Kubi concession, Ghana. Hydrothermal quartz is usually formed from magma-induced heat-related anomalies that are active in oceanic and continental crusts because of the redistribution of energy and mass in water-flowing environments [3, 4]. In hydrothermal deposits, quartz is seen to contain compositions of trace elements in large quantities as compared to other minerals [4]. This quartz is part of silicate-rich minerals in magma or ore-forming fluids which cools and crystallizes out of solution to form a denser conglomerate of minerals [5]. These hydrothermal conditions cause quartz crystals to form under certain growth conditions with peculiar textures, morphology, chemical, and physical properties. At higher temperature, quartz undergoes two mechanisms (i) dislocation creep which causes the quartz to assuage with increasing temperature and leads to a reduction in internal stress due to dislocation in the crystal lattice and (ii) migration of a series of microstructures of present grain boundaries by f accumulated strain energy between adjacent grains. These perceptible structures in quartz form continuous dynamic recrystallization textures or surfaces at different temperatures. The lowest temperature texture occurs at 250-400 °C [6] along grain boundaries and it is characterized by lumps or swells and small recrystallized grains. As the temperature is increased, the softening of the quartz also increases to cause a reduction in internal stress such that the recrystallized grains have no intergranular deformation characteristics. At temperatures between 500-550°C, the migrations mechanism along grain boundaries increases causing quartz to exhibit much larger proportions of recrystallized grains without any traces of original grains. Quartz is believed to be completely recrystallized at this temperature with much free grain boundaries devoid of any intragranular deformation characteristics but may contain or introduce other mineral grains as impurities.

Hydrothermal quartz has been studied by considering the quantity of water present based on OH contents with crystals exhibiting internal zoning with OH that varies from low to high defect levels [7] depending on the morphology or phase of quartz (smoky quartz possesses higher OH contents) [8]. Lin *et al.* [9], studied impurities in hydrothermal quartz using optical microscopy, electron probe microanalysis, scanning electron microscopy, inductively coupled plasma-optical emission spectrometry, and inductively coupled plasma mass spectrometry to explain the geological occurrence of a quartz deposit, mineralogical studies, and the processing technologies of hydrothermal quartz. These defects in quartz and related minerals introduce new properties into the material's structure and provide information on the physico-chemical conditions as well as the evolution and recycling of rock systems to quantify geological events.

There has not been much information on the characterization of hydrothermal quartz-vein hosting gold in Ghana to classify the quartz types, transformation, defects, and bonding. The sparse reports available have concentrated on geological activities at different depths; from mesothermal to hydrothermal quartz-vein gold mineralization [10, 11]. Tanner *et al*. [12], used cathodoluminescence, qualitative electron microprobe maps, and laser ablation inductively coupled plasma mass spectrometry on isotopic in situ quartz samples from USA and Australia and concluded that trace element signatures of quartz is an ineffective pathfinder in metal mineralization and the deposition of quartz occurred under a non-equilibrium process [12]. In our previous work [13], we attributed certain trace elements to indicator minerals such as quartz, magnetite, marcasite, and garnet from the Kubi Gold project of the Asante Gold corporation in Ghana.

Considering this, the identification and analysis of $SiO_2$-containing trace elements are crucial for exploring the hydrothermal conditions necessary for mineralization in $SiO_2$. This study employs advanced techniques including X-ray diffraction (XRD), air annealing furnace (AAF) measurements (thermal treatment process), transmission electron microscopy (TEM), energy dispersive X-ray spectroscopy (EDX), and differential scanning calorimetry (DSC). These methods are utilized to investigate the textural features, geochemistry (alteration and mineralization) of quartz minerals, phase diagrams, and defects in ore-related quartz from the Kubi Gold site in Ghana. The objective is to evaluate the variability of mineral concentrations in quartz formed under different flow regimes in hydrothermal systems.

## 2. Experimental details

### 2.1 Description of the field site

Figure 1 shows a geological map of the Kubi Gold Project near Dunkwa-On-Offin in the central region of Ghana where the samples were collected. In the Kubi Gold Project, the relief is distinguished by two linear southwest-northeast striking ridges. The western part of the concession follows the Ashanti trend [14] with steep to moderate slopes of about 30 degrees, stretching to a 300 m distance. The deposit of the Kubi Main area is overlapped by two major topographic features: a 50 m high initial mining site by AngloGold Ashanti and a 30 degrees steep slope hill with about 150 m elevation. Gold mineralization in Kubi occurs in about 15.0 meters thick garnetiferous horizon within the Birimian metasediment of the north-northeast trending shear zone of Birimian-Tarkwain contact [15]. The garnetiferous zone consists of fine-grained Au associated with indicator minerals such as pyrite, arsenopyrite, and pyrrhotite whereas the coarse-grained Au is concentrated in narrow veins of quartz. Within the oxides (alluvia zone) are highly altered limonite and hematite alterations. Apart from the main garnet zone, Au mineralization also occurs in quartz veins, veinlets, some stockworks, and the overall deposit trends northeast-southwest direction of the Birimian-Tarkwaian thrust and a trending basement fault along the north-south [14, 15]. The structure consists of foliations that strike at about 20 degrees and dip steeply along the east as the mineralization dips towards the west from the Birimian-Tarkwaian contact: characterized by minerals such as garnet, pyrite, magnetite, amphibole, pyrrhotite, and free gold within quartz veins [14]. This makes quartz veins in Kubi possess preferred orientations of shear fractures and other major fractures in a brittle-ductile shear zone. These quartz veins are seen to crosscut the garnet-rich zone into the parent or host rock. In view of this, high-grade flares within the mass (rock unit) consisting of auriferous and cross-cutting veins are worth investigating.

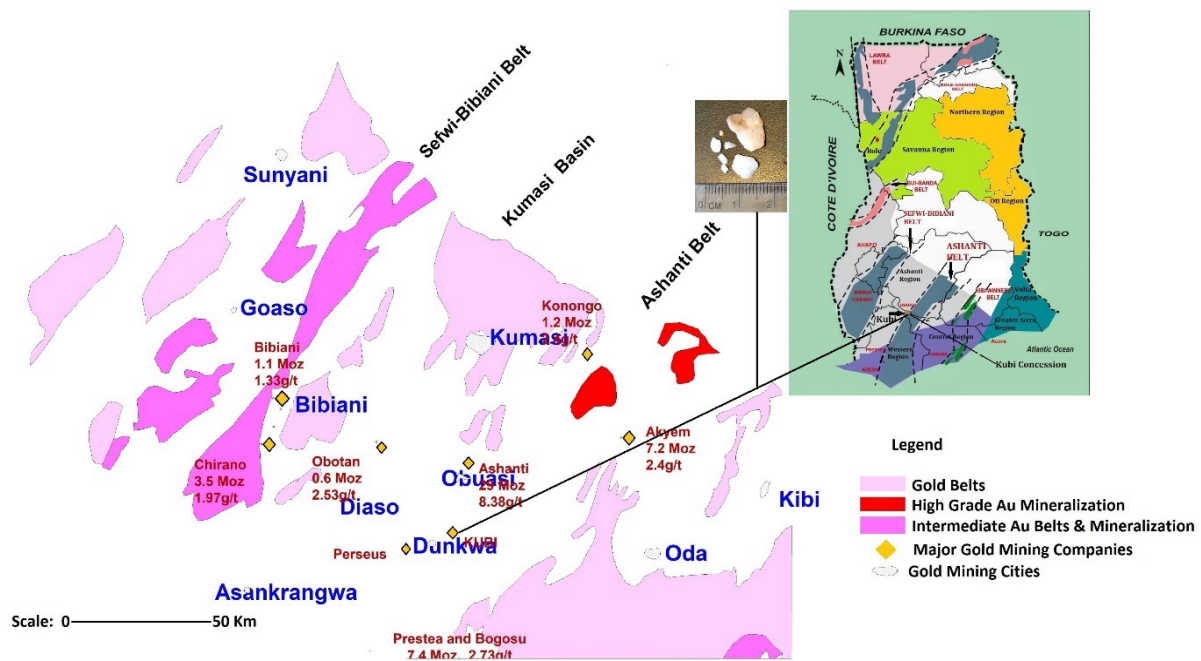

**Figure 1:** Geological map of mining areas in and around Dunkwa-On-Offin in the central region of Ghana and hand specimen samples of the quartz. [Map generated by G. K. Nzulu from MapInfo software]

### 2.2 Sample selection and preparation

Four different quartz ($SiO_2$) samples of light brown powder (outcrop sample as a reference sample taken with the aid of a geological hammer) containing little iron stains (Fig. 2a), light grey powder (artisanal pit sample of 5 m deep) with Fe stains as impurities and altered (Fig. 2d), black and white powder (small scale mining sample from a 12 m depth) containing other intrusive (Fig 2g), and pyrite veining quartz powder (drill core sample of about 30 m deep) containing other minerals (Fig. 2j) were sampled at random from an artisanal mining site at the Kubi Gold Project near Dunkwa-On-Offin. Each sample was crushed and sieved into a fine-grained powder sample and the four samples of $SiO_2$ origin were examined by X-ray diffraction. Thereafter, the samples were divided into two parts. One part of each sample was annealed at a temperature of 1100°C using an air furnace annealing chamber to analyze the effect of temperature on the impurity minerals and possible trace elements. This was done to examine the effect of temperature on impurity minerals in terms of decomposition, sublimation, precipitation, oxidation, reduction, and alteration after rock minerals/materials have been formed or protruded from similar conditions. This annealing was also performed to examine the effect on mineral grains, facilitation of the water-heat transfer of ions between and within minerals, and any growth of new minerals.

X-ray diffraction analysis was performed after annealing the first half of the samples. The other part of the samples was subjected to DSC scanning to analyze the structural transformation of the quartz phases.

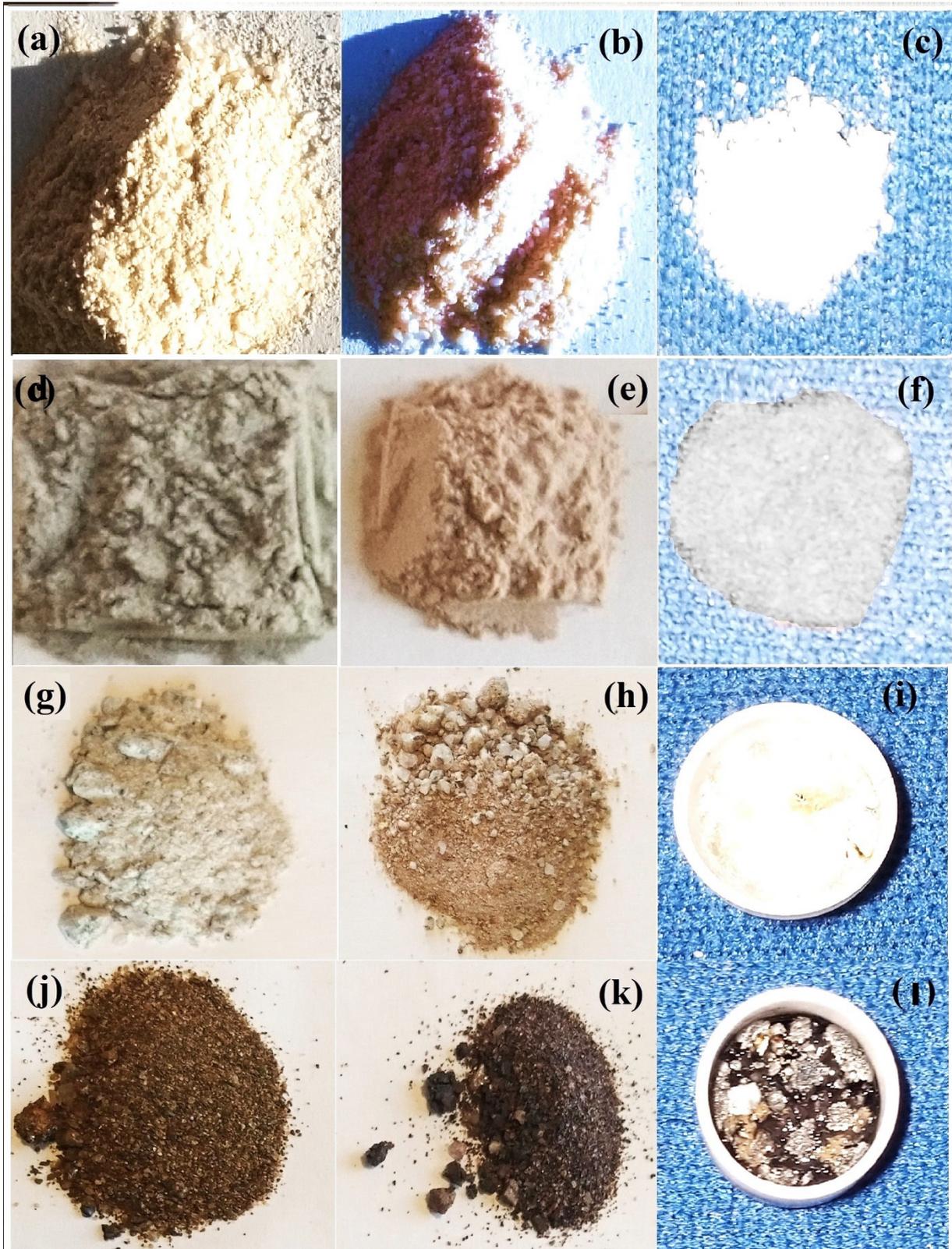

**Figure 2**: (a) fresh light brown powder quartz, (b) light brown powder quartz after annealing, (c) light brown powder quartz after DSC scan, (d) fresh light grey powder quartz, (e) light grey powder quartz after annealing, (f) light grey powder quartz after DSC scan (g) fresh ash-grey powder quartz, (h) ash-grey powder quartz after annealing, (i) ash-grey powder quartz after DSC scan, (j) fresh pyrite vein powder quartz, (k) pyrite vein powder quartz after annealing, and (l) pyrite vein powder quartz after DSC scan. [Photos by G. Nzulu].

## 2.3 Experimental details

The mineralogical compositions of the samples were primarily investigated by XRD diffraction. The XRD analysis was performed using a PANAnalyical X'pert [16] powder diffractometer at IFM, Linköping University, with a theta-2theta configuration and Cu-Kα radiation wavelength of 1.5406 Å. The Cu long-fine-focus tube was set to 45 kV and 40 mA, the scan step size of 0.0084° with a counting time of 19.68 s per step, and the scan range between 10 and 100 degrees in 2-theta scans. With the aid of the MAUD software [17] the data were quantitatively analyzed using Rietveld refinement where analytical functions of pseudo-Voigt and Pearson VII and Marquardt Least Squares algorithm [18] were used to model the experimental XRD patterns. The fitting parameters (background, peak shape, crystal structure, microstructure, strain, and texture) were used for the refinement while considering the least squares for the iteration [17] to minimize the residual parameter as well as fitting the anisotropic size-strain broadening for the crystallite size and possible micro-strain [19], and the background slope from increasing x-ray scattering at low angles was excluded in the final plot. The quartz samples were subjected to annealing using a compact 2" tube furnace of OTF-1200X-S-DVD type from MTI corporation [20]. The furnace has an inner sample sliding glass tube, where the samples were manually placed for rapid thermal processing (RTP) under room atmosphere conditions for two hours and at a furnace temperature of 1100 °C which is above the recrystallization temperature of quartz (500-550 °C). The four quartz samples were analyzed by HR-TEM, and selected area electron diffraction (SAED) by using a Tecnai G2 20 U-Twin 200 kV FEGTEM microscope [21]. Differential scanning calorimetry (DSC) experiments were conducted on quartz powder samples using a Netzsch STA 449 C instrument. The samples were placed in alumina crucibles and heated from room temperature to 1400 °C at a rate of 20 °C/min under an argon flow atmosphere. The DSC curves were processed using the Proteus software [22].

## 3. Results

### 3.1 XRD results

Figures 3-6 show X-ray diffractograms of the quartz powder samples (fresh and annealed) displayed in Fig. 2 (a - h) with the results of Rietveld refinement assuming pure $SiO_2$ including the residual of the fit [17, 23]. The ten pronounced peaks for the fresh light brown quartz, eleven distinct peaks of the light grey quartz, nine well-defined peaks from the ash-grey quartz, and sixteen clearly observed peaks from the pyrite veining quartz were indexed. The lattice parameters were found to be $a$ = 4.92 Å and $c$ = 5.42 Å of the trigonal structure with space group $P3_221$. These peaks are in agreement with reference data [24, 25] and consistent with literature values [23]. The diffractograms of the annealed quartz samples showed additional well-pronounced peaks of hexagonal $SiO_2$ structure ($P6_222$ space group) with lattice parameters $a$ = 5.015 Å and $c$ = 5.471 Å, which hold true for the structure of β-quartz [26] and in agreement with reference data [24, 25] for $SiO_2$.

The diffractogram in Fig. 5(a) reveals five diffraction peaks that are indexed as biotite with lattice parameters $a$ = 5.247 Å, $b$ = 8.992 Å, and $c$ = 10.091 Å of space group C12/m1 corresponding to the monoclinic crystal structure. These peaks agree with reference data [27] and are consistent with the literature [28]. From the diffractogram of the pyrite veining quartz in Fig. 6 (a) we found other mineral phases of which the most well-defined phase is indexed as cubic pyrite with lattice constant $a$ = 5.417 Å and of space group Pa-3. This is in agreement with reference data [29] and consistent with literature values [30]. The next set of observed peaks in the diffractogram is indexed as orthorhombic magnetite of lattice parameters $a$ = 9.273 Å, $b$ = 9.239 Å, and $c$ = 2.746 Å with space group Bbmm and in accordance with reference data [31] and consistent with literature values [32]. Also observed from the diffractogram are weak peaks of an impurity mineral phase indexed as monoclinic titanium oxide with lattice parameters $a$ = 5.8555 Å, $b$ = 9.340 Å, and $c$ = 4.142 Å and of space group A112/m that agrees with reference data [33].

The diffractograms for all annealed quartz samples in Figs. 3b, 4b, 5b, and 6b reveal peaks that are indexed as cubic hematite of lattice parameter $a$ = 5.43 Å and of space group R-3c; consistent with the literature [34]. Table I, lists the refined values of all parameters of the quartz samples, and appendix A and B give the structural refinement parameters of all powder quartz samples with their impurity minerals for both pre-and post-annealing conditions, respectively.

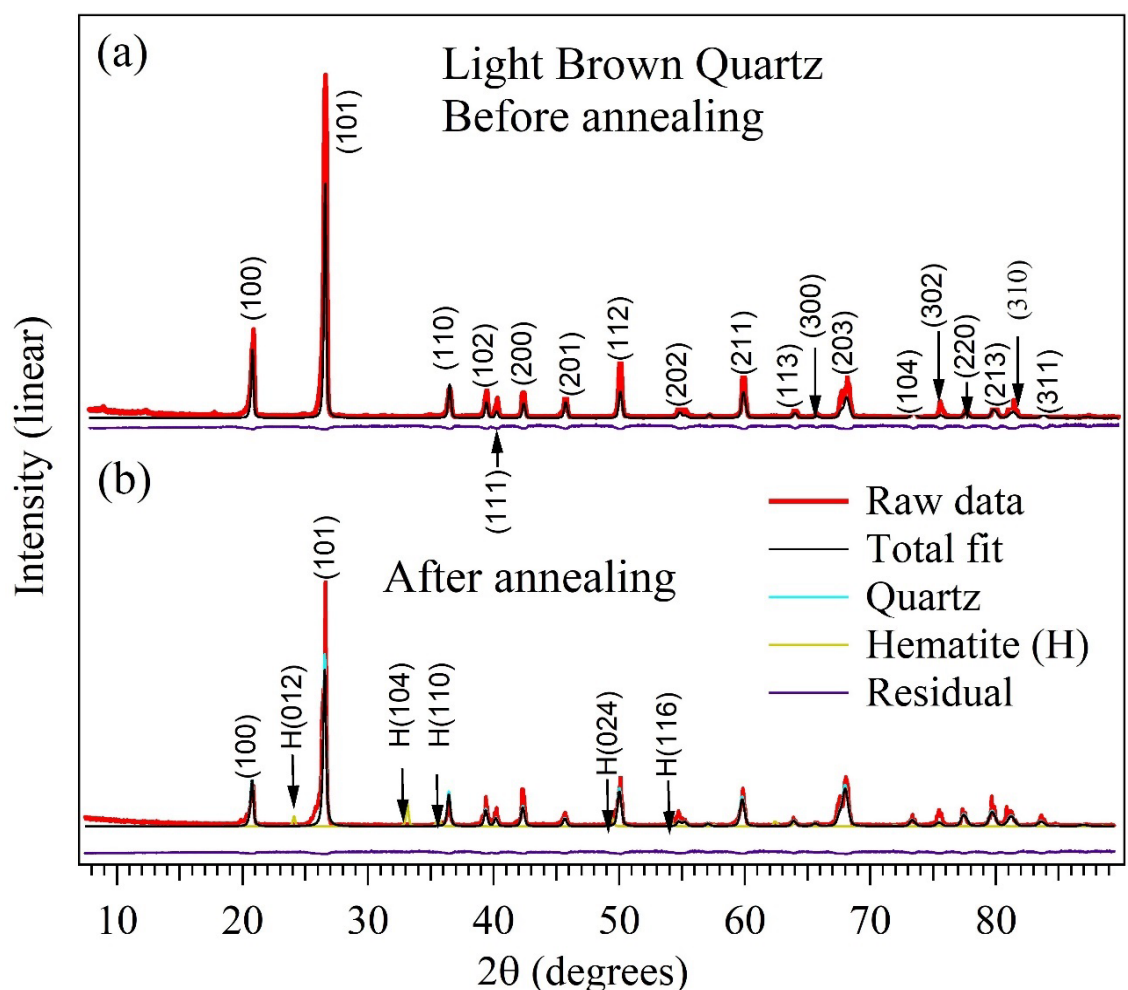

**Figure 3:** X-ray diffractogram of the (a) light brown powder quartz sample with Fe stains and (b) annealed to 1100 °C, showing distinct peaks indexed by quartz and hematite.

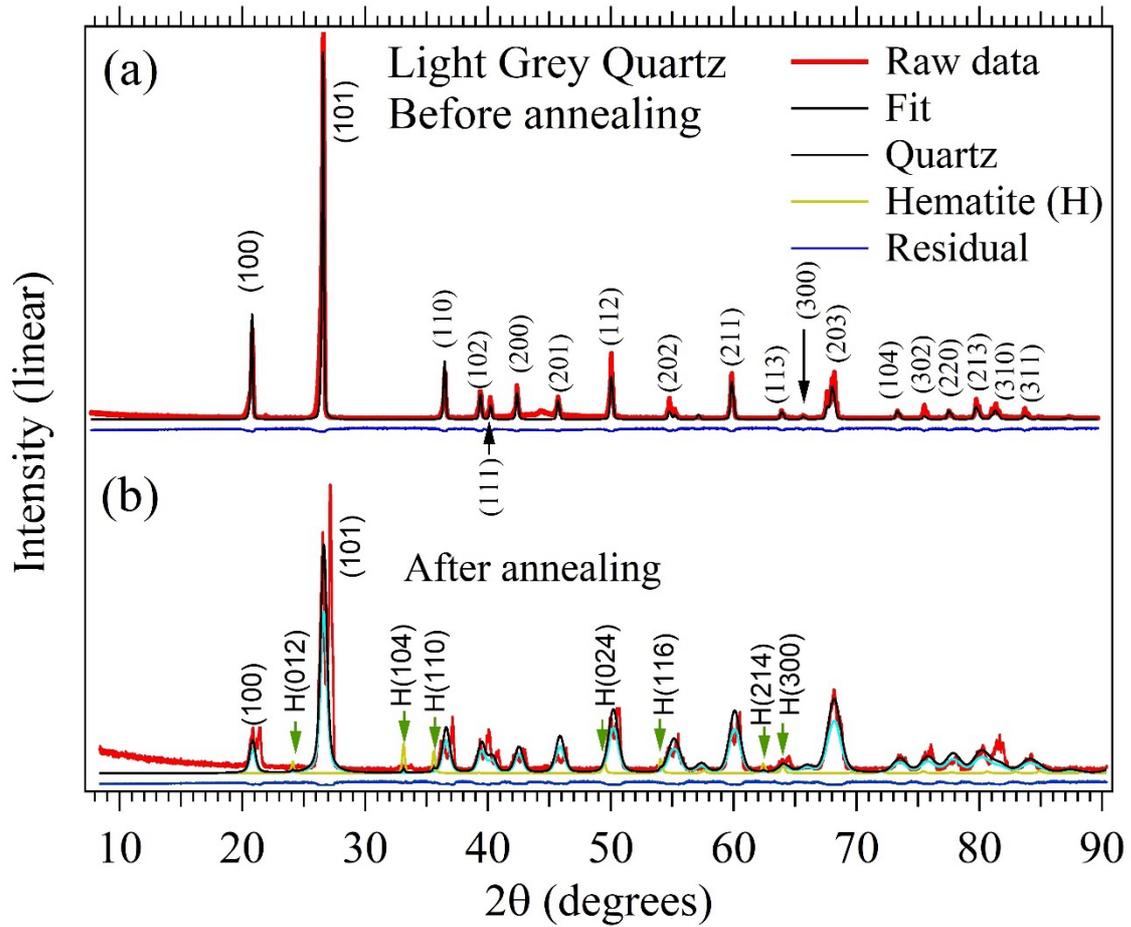

**Figure 4:** X-ray diffractogram of (a) light grey powder sample with Fe stains and (b) annealed to 1100 °C, indexed by quartz and hematite.

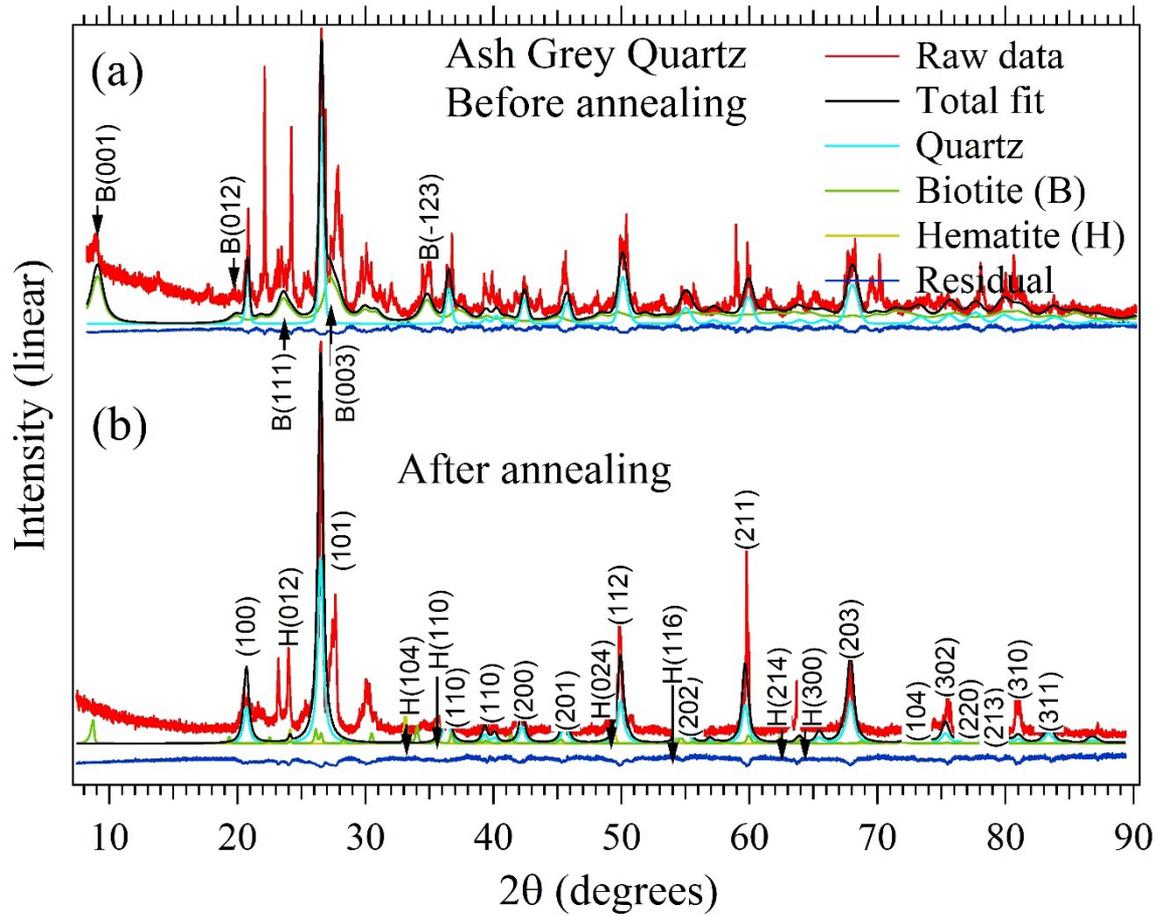

**Figure 5:** X-ray diffractogram of (a) ash-grey quartz sample and (b) annealed to 1100 °C, with peaks indexed by quartz, biotite, and hematite minerals.

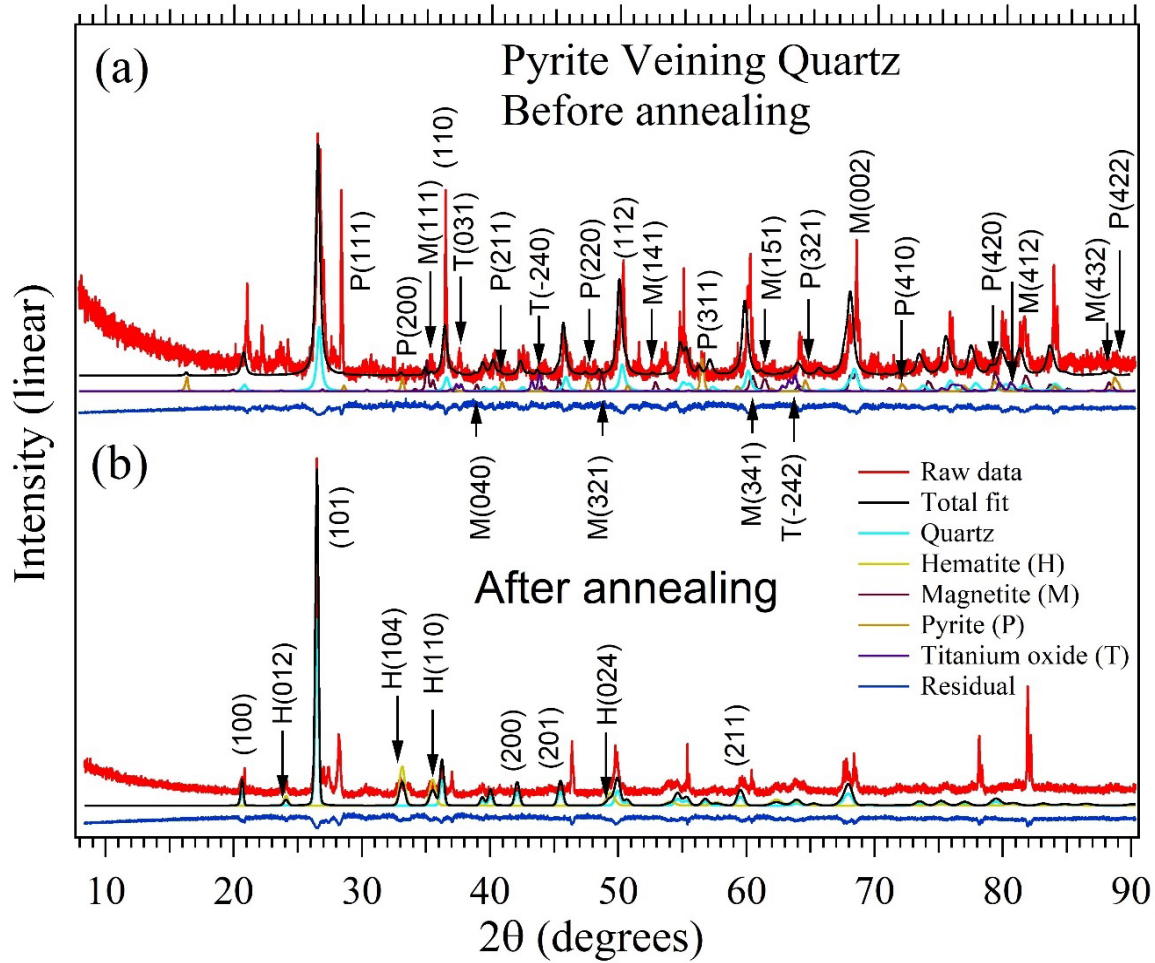

**Figure 6:** X-ray diffractogram of (a) pyrite veining quartz powder sample and (b) annealed to 1100 °C, with peaks indexed by quartz, hematite, magnetite, pyrite, and titanium oxide minerals.

### 3.2 TEM results

Figure 7 (a-h) shows TEM patterns of all four quartz powder samples where the light brown in Fig. 7a & b and light grey quartz in Fig. 7c & d patterns reveal only Si and O elements in the pure $SiO_2$ structure with little (Fe stains) or no impurities. In contrast, many other phases were found together with quartz in the ash-grey and pyrite vein samples as shown in Fig. 7e-h. The EDX spectra obtained from the low-contrast phase showed strong K X-ray peaks for Si and O for all quartz samples and additional elements of Fe, Na, Al, Ca, Mg, Ti, and S in the ash-grey and pyrite vein quartz. The elemental distribution in the ash-grey quartz sample: Si, O, Mg, Fe, Al, K, Na, and Ca, suggest that the sample may contain minerals such as potassium feldspar – $K(AlSi_3O_8)$, biotite ($H_4K_2Mg_6Al_2Si6O_{24}$), and $CaCO_3$. However, the refinement in MAUD favored the presence of biotite as the main contributing impurity phase mineral. With respect to the pyrite vein quartz, the elemental distributions of Si, O, Fe, S, Ti, and Al suggest the presence of pyrite, titanium oxide, magnetite, or other Fe-silicate minerals. These elements are components of minerals that have been observed and fitted by the XRD measurements and are contained as impurities in the host rock [29-34]. Table II gives quantitative results for the assigned peaks in each sample obtained from EDX at different focused spots on the samples. The weakly distributed impurity elemental components produced characteristic peaks in the EDX spectra indicating that impurity phase minerals could be present in the samples.

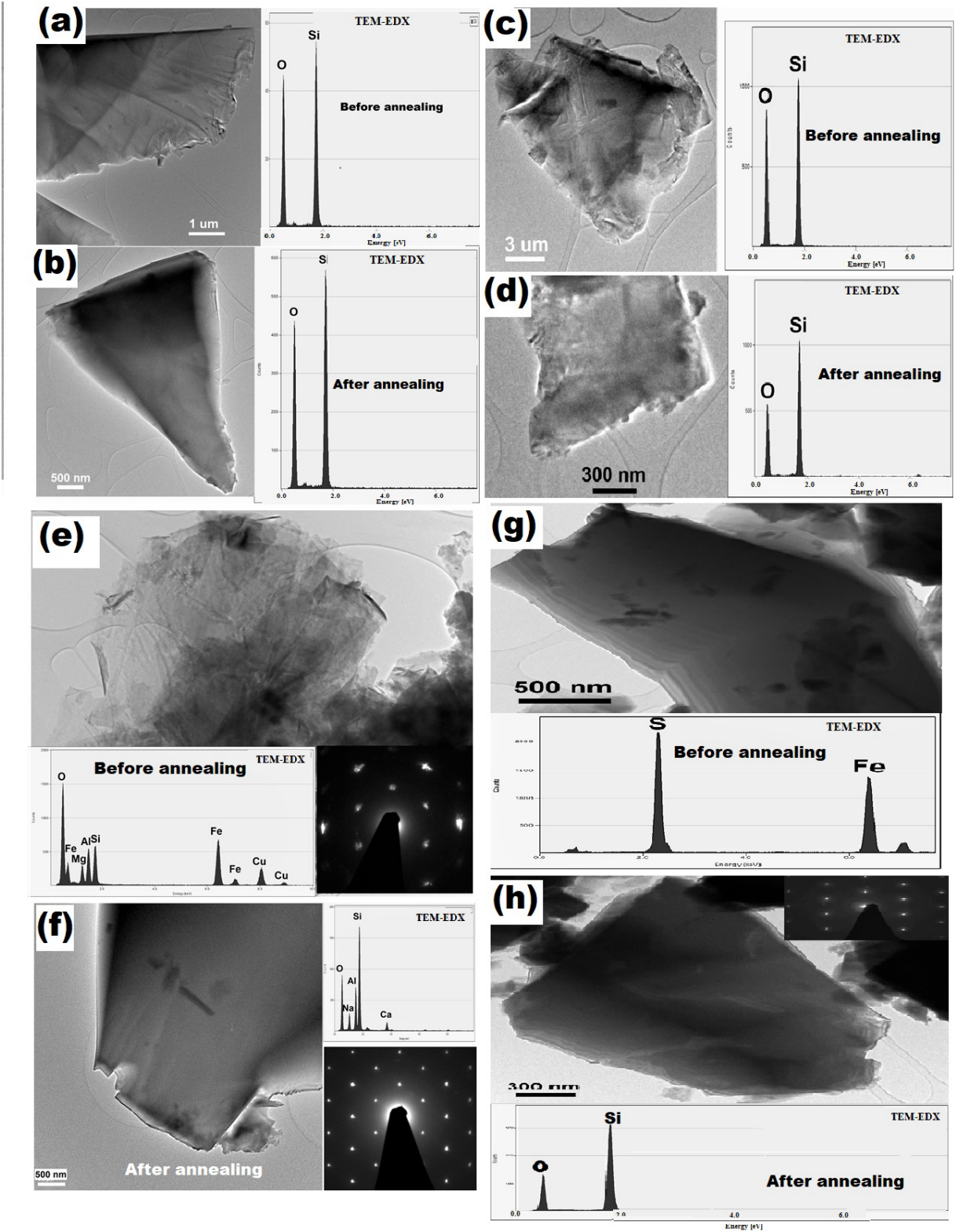

**Figure 7:** TEM images, SAED patterns, and EDX spectra showing the elemental compositions in (a) light brown $SiO_2$ before annealing, (b) light brown $SiO_2$ after annealing, (c) light grey

SiO₂ before annealing, (d) light grey SiO₂ after annealing, (e) ash-grey SiO₂ before annealing, (f) ash-grey SiO₂ after annealing (g) pyrite vein SiO₂ before annealing and (h) pyrite vein SiO₂ after annealing.

## 3.3 DSC results

Figure 8 (a-d) shows DSC curves for fresh (pre-annealed) quartz samples displaying the structural transition from the α-quartz phase to β- quartz and finally to the β-cristobalite modification.

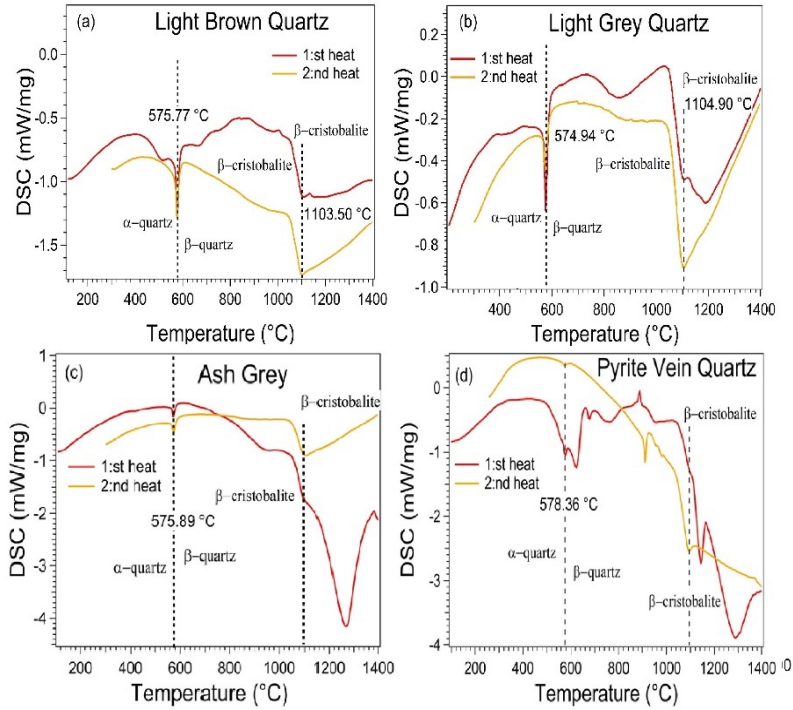

**Figure 8:** DSC curves for all quartz samples showing the transition between α, β quartz, and finally to the β-cristobalite phases.

Upon heating the samples to 1400 °C, a sharp endothermic peak appears in the 574-579 °C temperature range with the deepest dip at 575 °C for the quartz samples that is reversible during cooling. From 1000-1200 °C, another endothermic reaction occurs in all samples that is reversible during cooling under an exothermic process. The temperatures at which the transitions occur are in agreement with the reference data [35]. The successive cooling data recorded from 1400 °C back to room temperature (for all samples) were not plotted as they appeared to be very weak and less sharp endothermic peaks due to impurity phases. These impurity phases show a transformation between 600 °C and 725 °C with the highest rate at 680 °C alongside the β-quartz phase in the pyrite vein powder quartz sample in the first and second heating cycles as observed in Fig. 8d. This could be a titanium monoxide phase which is also observed around 800 °C and in agreement with the literature [36]. The pyrite in the quartz powder sample first transforms into iron (III) sulfide before dissociating into hematite [37]. Below 400 °C the magnetite in pyrite vein quartz powder samples may undergo a transition from magnetite to maghemite and from maghemite to hematite at higher temperatures. This could be related to the large surface-to-bulk ratio of the fine powder samples [38]. These are possible conditions to alter the quartz samples and influence the successive cooling peaks.

# 4. Discussion

## 4.1 Temperature dependent changes in microstructure and recrystallization

From the combined investigations in this study, we examine the effect of temperature on impurity minerals in quartz in terms of decomposition, oxidation, reduction, and alteration. While the thermal history of the as-collected outcrop quartz samples is unknown, the present annealing shows significant changes to the microstructure, thus indicating that either the samples were formed at relatively low temperature or had somehow transformed over geological time at reduced temperature compared to any high-temperature formation process. From the XRD results in Figs. 3, 4, and 5 it can be observed that the diffraction peaks of the annealed samples are sharper and very well oriented while the fresh quartz samples have more intense peaks. The diffractogram in Fig. 6 appears different as the observed peaks of the fresh samples have sharper peaks due to a higher preferred orientation than the annealed quartz samples. This can be attributed to the recrystallization and recovery of crystal grains and structures that can be associated with oxygen vacancies and cationic contributions from other impurity minerals iron disulfide, magnetite, and titanium oxide [39,40]. Thus, quartz undergoes dislocation creep and grain boundary migration of microstructure as temperature increases to cause an increase in strain energy. At this discernible stage, an active recrystallization may occur with textures of minerals of iron disulfide, magnetite, and titanium oxide at elevated temperatures.

**Table 1:** Refined lattice parameters of four quartz samples before and after annealing. $R_p$ quantifies the residual difference between the observed and calculated data points on a point-to-point basis. $R_{wp}$ is the weighted profile of the residual that weighs the higher-intensity data points over the lower-intensity data points. $R_{exp}$ is the expected best possible residual factor.

| Quartz type | Condition | Lattice Parameter | Volume of unit cell ($Å^3$) | Rietveld Refine Parameter (%) |
|---|---|---|---|---|
| Light brown | Fresh | $a = b = 4.919$ Å, $c = 5.409$ Å, $α= β= 90, γ= 120$ | 113.23 | $R_p = 12.65$, $R_{wp} = 23.61$, $R_{exp} = 7.93$ |
| | Annealed | $a = b = 4.926$ Å, $c = 5.416$ Å $α= β= 90, γ= 120$ | 113.811 | $R_p = 32.63$, $R_{wp} = 13.82$, $R_{exp} = 3.75$ |
| Light grey | Fresh | $a = b = 4.923$ Å, $c = 5.416$ Å $α= β= 90, γ= 120$ | 113.67 | $R_p = 16.19$, $R_{wp} =28.05$, $R_{exp} = 7.46$ |
| | Annealed | $a = b = 4.945$ Å, $c = 5.440$ Å $α= β= 90, γ= 120$ | 115.20 | $R_p = 18.32$, $R_{wp} =23.90$, $R_{exp} = 3.50$ |
| Ash grey | Fresh | $a = b = 4.923$ Å, $c = 5.413$ Å $α= β= 90, γ= 120$ | 113.53 | $R_p = 14.23$, $R_{wp} = 26.28$, $R_{exp} = 9.27$ |
| | Annealed | $a = b = 4.934$ Å, $c = 5.426$ Å $α= β= 90, γ= 120$ | 114.39 | $R_p = 9.99$, $R_{wp} = 13.04$, $R_{exp} = 4.71$ |
| Pyrite vein | Fresh | $a = b = 4.923$ Å, $c = 5.406$ Å, $α= β= 90, γ= 120$ | 113.46 | $R_p = 23.26$, $R_{wp} = 28.97$, $R_{exp} = 8.90$ |
| | Annealed | $a = b = 4.943$ Å, $c = 5.419$ Å $α= β= 90, γ= 120$ | 114.66 | $R_p = 15.46$, $R_{wp} =14.75$, $R_{exp} = 6.89$ |

The results in Table I show that the lattice parameters of the annealed quartz samples gradually increase for the samples including those with little or insignificant impurity levels as the effect of temperature on the air-annealed quartz samples is observed by changes in lattice constants. The changes in lattice constants can be attributed to oxygen vacancy concentration, and the removal or substitution of cations or alkali metals in the vicinity of the vacancies [36].

**4.2 Devitrification and isomorphic substitution processes**
It should be noted that the Kubi Gold mining concession is an old open pit mine that has undergone a reclamation process (nature restoration), hence the materials have experienced *e.g.*, sedimentation, crystallization, metamorphism, and alterations. The presence of foreign materials in some of the quartz minerals is an indication that those quartz minerals might have experienced a rearrangement process known as *devitrification* [41] to re-order the atoms before the material cooled to the solid state. The devitrification process can be caused by contamination in the quartz through the introduction of alkali metals at elevated temperatures above 1000 °C [41, 42]. In the hydrothermal system, contamination of quartz may also occur in the presence of water (droplets) as the quartz extrudes from the surface. Devitrification is enhanced by the atmosphere in the presence of oxygen and activated by temperature. Thus, the higher the temperature the faster goes the quartz devitrification process [41]. During the hydrothermal flow regime, water, and trace elements experience a reduction process with the incorporation of OH, which is influenced by increasing temperature to enhance the solubility of trace metals and by substitution of both mono and trivalent cations in the hydrothermal fluid [8]. These hydrothermal quartzes are heterogeneous in nature with the introduction of OH and this heterogeneous behavior is concentrated at the center of the mineral [43], implying that water is an important fluid inclusion mintage [44] in the hydrothermal system and very rich in alkali metals. Given this, the preservation of OH defects in quartz in the hydrothermal system may assist geologists to estimate the time scale of volcanic activities.

The minerals biotite, pyrite, magnetite, and titanium oxide that act as impurities in the ash-grey and pyrite-veining quartz samples may have been deposited through pre-existing fractures or fissures when the quartz rock was filled with new mineral materials in the hydrothermal flow regime. Minerals and trace elements can also be introduced into quartz by different mechanisms such as isomorphic substitution and valence balancing reactions [45]. The chemical bond in quartz between the Si and the O atoms enables the $Si^{4+}$ ions to easily be substituted by other elements with similar ionic radius, charge, and valency [45]. This makes it possible for tetravalent elements such as $Ti^{4+}$ to occupy the $Si^{4+}$ lattice position of the Si-O tetrahedron by isomorphic substitution (incorporation of tetravalent elements into the crystal lattice) [46]. Therefore, the Ti in the pyrite vein powder quartz sample occupies the $Si^{4+}$ lattice position of the Si-O tetrahedron by isomorphic substitution. In our previous work [13], we identified Si, Fe, Al, Ti, Hg, and S as the main possible pathfinder elements of Au in alluvia sediments in the same sample location with negligible contributions from Na, Li, and K. These elemental components are probably widely spread across the entire concession and can be associated with the Au indicator minerals. Based on charge neutrality considerations and valence balancing [2, 47], the inclusion of interstices in the quartz crystal lattice is possible with the trivalent elements $Fe^{3+}$ and $Al^{3+}$ in this concession area and present in the ash-grey and pyrite vein quartz, with some alkali elements $Li^+$, $K^+$, $Na^+$, and $H^+$ [46]. Other conditions favor the incursion of trace elements and metals in the growth of quartz in hydrothermal systems that includes pH, temperature, pressure, fluid chemical composition, and rate of quartz crystal growth [2]. These conditions make it very feasible for most trace elements to be substituted into the lattice position of $Si^{4+}$. These general conditions for hydrothermally grown quartz and other minerals according to Mumin *et al.*, [11] are favorable in the Ashanti Gold belt of Ghana that cut across the Kubi gold concession.

**4.3 Influence impurities and precipitation**
Natural occurring glassy quartz is somewhat fragile, possesses many defects such as point defects, line defects, lattice defects, interstitial defects, and three-dimensional defects, and can easily be contaminated during precipitation and crystallization of the fluid (minerals) in hydrothermal systems [48]. This implies that the elements Fe, Mg, Na, K, Al, Ti, S, and Ca that are contained in indicator minerals at the Kubi concession such as the pyrite and silicate group minerals can easily be incorporated into the Si-O tetrahedron as contaminations in the ash-grey and pyrite vein quartz samples. These contaminants explain changes in different geological processes such as primary crystallization, metamorphism, the difference in crystallization temperatures, mineral alterations, and secondary dissolution-precipitation [49, 50]. Thus, the presence of indicator minerals, metals, and other trace elements in these quartz samples are indications that during the formation of the ore in the hydrothermal system, quartz crystals experienced precipitation under certain physico-chemical conditions at different crystallization processes. The presence of metal complexes in the materials is attributed to changes in fluid conditions and reactions of fluidic minerals that caused materials to precipitate in the form of sulfides and other indicator minerals within the quartz crystals and these materials are reserved in the ore-forming fluid during the flow regime.

From the annealed quartz samples in Fig. 2 (b, d, f, and h), we observed color changes in the samples from their original state to orange-reddish brown (present states of samples) and this may be attributed to the state of oxidation with temperature acting as a catalyst. The light brown and grey fresh quartz samples that showed pure $SiO_2$ with insignificant impurities including possible iron stains experienced oxidation to induce hematite alterations in the quartz minerals when annealed. The biotite in the ash-grey quartz sample may decrease with increasing temperature [51] and break down into other constituents at an elevated temperature of 900 °C. As the sample is annealed, these constituent elements either sublime or decompose, and the remaining parts oxidize to form other impurities that were below the detection limit of the XRD instrument.

The minute TiO which is common in most natural quartz and present in the pyrite vein quartz sample releases $Ti^{4+}$ which is thermodynamically reactive and burns at low temperatures in normal atmospheric conditions [52]. The Ti, if present in this pyrite-vein quartz, is below the detection limit of the X-rays and at high temperature diffuses or decomposes during the annealing and is not apparent to cause the quartz to become rutilated.

From the diffractogram of Fig. 6 (a), the pyrite observed in the pyrite veining quartz experienced oxidation when annealed in the air to convert sulfur into sulfate and ferrous iron including other elemental sulfur products such as polysulfides, hydrogen sulfide, ferric hydroxide, and iron oxides (hematite) [53]. It is also possible for the sulfur to sublime during the annealing process leaving Fe atoms which then react with oxygen atoms to form hematite [34]. The magnetite in the pyrite vein sample also experienced oxidation during annealing (oxygen fugacity) to produce hematite according to the reaction 4 $Fe_3O_4$ (magnetite) + $O_2$ → 6 $Fe_2O_3$ (hematite) [53]. These minerals and elemental particles either decomposed, sublimed, or experienced oxidation and reduction processes during annealing to form other impurity phases and alterations (hematite). The monoclinic biotite mineral structure in the ash-grey quartz contains hexagonal crystal planes of $SiO_2$ tetrahedrons with high mechanical strength that is bonded in layers by cations ($K^+$, $Mg^{2+}$, $Ca^{2+}$, and $Al^{3+}$) to attract available valences of $SiO_2$ tetrahedrons. As observed in the ash-grey samples, water particles break into the biotite minerals to release the bonding ions or elements [50] and similar phenomena occur for titanium monoxide and the oxides in the pyrite-vein quartz sample during the flow regime in the hydrothermal system.

**Table II:** Quantitative analysis results from EDX for the four quartz samples before and after annealing.

| Quartz type | Condition | Elements | Atomic percentage at.% |
|---|---|---|---|
| Light brown | Fresh | Si | 30.00 |
|  |  | O | 70.00 |
| Light brown | Annealed | Si | 33.00 |
|  |  | O | 67.00 |
| Light grey | Fresh | Si | 30.00 |
|  |  | O | 70.00 |
| Light grey | Annealed | Si | 32.00 |
|  |  | O | 68.00 |
| Ash- grey | Fresh | Si | 27.00 |
|  |  | O | 44.00 |
|  |  | Mg | 4.00 |
|  |  | Fe | 7.00 |
|  |  | Al | 5.00 |
|  |  | K | 6.00 |
|  |  | Na | 6.00 |
|  |  | Ca | 1.00 |
| Ash- grey | Annealed | Si | 31.00 |
|  |  | O | 43.00 |
|  |  | Fe | 5.00 |
|  |  | Na | 5.00 |
|  |  | Ca | 1.00 |
|  |  | Al | 6.00 |
|  |  | Mg | 3.00 |
|  |  | K | 5.00 |
| Pyrite vein | Fresh | Si | 25.00 |
|  |  | O | 40.00 |
|  |  | Fe | 11.00 |
|  |  | S | 17.00 |
|  |  | Ti | 4.00 |
|  |  | Al | 3.00 |
| Pyrite vein | Annealed | Si | 29.00 |
|  |  | O | 52.00 |
|  |  | Fe | 17.00 |
|  |  | Al | 2.00 |

### 4.4 Endothermic and exothermic reactions in quartz

From the DSC results, the light brown and grey quartz in Fig. 8 a, b experienced almost the same heating mechanism during the first cycle which could be attributed to the presence of impurities or alterations. In the second heating cycle, we observe a smooth reaction as the impurities have decomposed or evaporated at elevated temperatures with endothermic reactions occurring at temperatures 575.77 °C, 574.94 °C, and 1103.50 °C, 1104.90 °C respectively with peak maxima. In the reverse reaction (cooling phase), a favorable exothermic reaction for the transition of cristobalite to quartz occurs for only the lightly altered (light brown and grey) quartz samples. In this DSC measurement, an endothermic reaction occurs with little or no oxygen to create some moist that led to a reduction mechanism which changed the color of the light brown quartz to white and light grey quartz to dark grey quartz as observed in Fig. 2 (c) and (f) respectively. The ash-grey and pyrite vein quartz also changed to creamy white and chocolate-dark-brown-white quartz as seen in Fig. 2 (i) and (l) due to the presence of other impurity minerals (biotite, pyrite, magnetite, and titanium dioxide).

In both air annealing and DSC measurements, there are transformations from one quartz phase to another. At room temperature and normal pressure quartz is stable as α-quartz with space group $P3_221$ for all four samples. Above 570 °C, the α-quartz transforms to hexagonal with space group $P6_222$ [54]. At this stage the quartz transforms to β-quartz and becomes stable around 880 °C, a reversible process influenced by symmetry loss and induced by a tilt of the tetrahedron along the *b-axis*. At elevated temperatures, the materials undergo a series of transformations to either orthorhombic α-tridymite, hexagonal β-tridymite, tetragonal α-cristobalite or isometric β-cristobalite.

The DSC curves in Fig. 8 show the transformation from α-quartz to β-quartz phase above 570 °C and cristobalite above 900 °C. The α-quartz, which is stable at room temperature before annealing, experienced a reversible change in the crystal structure to form β-quartz at higher temperatures and this is accompanied by an increase or expansion of crystal size as shown in Table I. The expansion in crystal size favors oxidation reactions to cause chemical effects by mineralogical alteration at higher temperatures where the present quartz samples are altered by hematite. At temperatures below 400 °C [6] texture growth may occur along grain boundaries with bulges and recrystallized grains. An increase in temperature softens the quartz materials or samples to release internal stress to cause mineral components or impurity minerals in addition to the quartz to undergo recrystallization without intergranular deformation characteristics. At temperatures around 550°C, there is an increase in the migrations of grain boundaries to promote larger fractions of quartz leading to a recrystallized material, and at elevated temperatures, the impurity minerals either decompose or undergo oxidation and reduction process leading to alterations on the quartz minerals.

These defects in quartz and alterations enable the mineral to act as a geochemical and petrological pointer, and as a material source for potential interpretation and analysis of sediments and sedimentary rocks.

It can be anticipated that the abundant quartz minerals at the Kubi mining area, from the near surface (outcrops) to greater depths have experienced similar physical, chemical, and geological conditions and protruded from a hydrothermal system where the minerals have been precipitated by hot fluids and altered by veins of surrounding rocks. Thus, during the transport process or flow regime, part of the host rocks dissolved and reacted chemically with the fluid to form mineral indicators (pyrite, magnetite, titanium monoxide, and biotite) or impurities (Fe, Mg, Na, K, Al, Ti, S, and Ca) in the quartz minerals as they get deposited and solidify at different stages.

## 5. Conclusions

The present study has revealed that temperature-dependent structural changes occur in impurity phases in quartz from hydrothermal deposits of vein types containing fractions of silicate sediments and other mineral species. This gangue quartz does not only act as an indicator of gold but also contains other indicator minerals of gold such as hematite, magnetite, titanium monoxide, and other sulfate minerals like pyrite, arsenopyrite, and chalcopyrite. The XRD results of the fresh and annealed quartz samples indicate that both pure and mineralized (impurity phase) quartz can experience hematite alteration at elevated temperatures under both oxidation and reduction processes or mechanisms. These transformations are also observed from DSC scans where quartz changes from α-quartz to β-quartz and to a more stable high temperature quartz (cristobalite). The results indicate that the largest amount of impurities in the quartz samples before annealing are biotite, pyrite, titanium monoxide, and magnetite in addition to some heavier elements, but these were present with very low content. The result from the air annealing shows that the process of decomposition of minerals (magnetite and pyrite) and elemental particles in the quartz in the oxide atmosphere depends on temperature, where the pyrite transforms to iron (III) sulfate and further oxidizes to hematite; and the magnetite to maghemite and further oxidizes to hematite, respectively, in the pyrite vein quartz sample. This temperature-dependent mechanism also facilitates the transformation of quartz by iron minerals, specifically the conversion of hematite to magnetite. XRD analysis before and after annealing the quartz samples reveals the incorporation or formation of new minerals within the silicon tetrahedron structure at elevated temperatures. The observed alterations of hematite are induced by thermal variations. This makes hematite and magnetite to serve as indicators for Au mineralization and alteration in the Kubi concession. The collected outcrop quartz samples had not been previously annealed in the ground. We conducted further annealing to demonstrate that the phase separation process, which had already been initiated by natural conditions, can be advanced or completed in the laboratory. The natural trends serve as valuable indicators, as our experimental results corroborate them. The elemental species in the samples such as Si, O, Ti, Fe, S, Mg, Na, Al, and Ca are identified by TEM and attributed to pure quartz and impurity minerals such as titanium oxide, biotite, pyrite as well as gangue minerals like magnetite and hematite (present as alteration minerals) in the samples acting as defects. The results indicate that annealing at high temperatures above 1000 °C, the mineral intrusions in the samples undergo phase transformations during diagenesis that produce devitrified $SiO_2$ crystallites to host impurity minerals to serve as good structural and mineralogical signatures. These signatures in quartz can be very useful in other geological settings to understand the mineralization and recrystallization mechanisms of quartz in other sediments.

It can be concluded that the stability of quartz and other minerals is influenced by temperature such that under high-temperature conditions, hematite-limonite alterations occur through oxidation and reduction processes, and these could be the contributing factor for alterations in many sediments within the Kubi concession area.

**Acknowledgements**

We thank the Swedish Government Strategic Research Area in Materials Science on Functional Materials at Linköping University (Faculty Grant SFO-Mat-LiU No. 2009 00971). M.M. acknowledges financial support from the Swedish Energy Research (Grant No. 43606-1) and the Carl Tryggers Foundation (CTS23:2746, CTS20:272, CTS16:303, CTS14:310).

## Appendix A

**Table (I):** Structural refinement parameters of light brown, light grey, ash-grey, and pyrite vein quartz powder samples before annealing
*************************************************************************

### i. Quartz

| | | |
|---|---|---|
| Symmetry: Trigonal | Space group = P3221 | COD ID: 9005017 |
| Wavelength CuKα = 1.5406 nm | | Ref. cell volume =113.010 Å³ |
| Wavelength CuKβ = 1.5444 nm | | Refined cell volume = 113.011 Å³ |

*************************************************************************

| OBSERVED | | | CALCULATED | | DIFFERENCE | |
|---|---|---|---|---|---|---|
| 2 Theta | d | h k l | 2 Theta | d | 2Theta | d |
| 20.788 | 4.2789 | 1 0 0 | 20.860 | 4.2550 | -0.072 | 0.0239 |
| 26.570 | 3.3595 | 1 0 1 | 26.640 | 3.3435 | -0.070 | 0.0160 |
| 36.438 | 2.4692 | 1 1 0 | 36.544 | 2.4569 | -0.106 | 0.0123 |
| 39.346 | 2.2932 | 1 0 2 | 39.465 | 2.2815 | -0.119 | 0.0117 |
| 40.140 | 2.2496 | 1 1 1 | 40.300 | 2.2362 | -0.160 | 0.0134 |
| 42.337 | 2.1378 | 2 0 0 | 42.450 | 2.1277 | -0.113 | 0.0101 |
| 45.654 | 1.9899 | 2 0 1 | 45.793 | 1.9799 | -0.139 | 0.0100 |
| 49.991 | 1.8270 | 1 1 2 | 50.139 | 1.8180 | -0.148 | 0.0090 |
| 54.670 | 1.6812 | 2 0 2 | 54.875 | 1.6717 | -0.205 | 0.0095 |
| 59.792 | 1.5489 | 2 1 1 | 59.960 | 1.5415 | -0.168 | 0.0074 |
| 63.861 | 1.4597 | 1 1 3 | 64.036 | 1.4529 | -0.175 | 0.0068 |
| 65.566 | 1.4258 | 3 0 0 | 65.786 | 1.4184 | -0.220 | 0.0074 |
| 67.972 | 1.3811 | 2 0 3 | 67.744 | 1.3821 | 0.228 | -0.0010 |
| 73.328 | 1.2929 | 1 0 4 | 73.468 | 1.2879 | -0.140 | 0.0050 |
| 75.409 | 1.2622 | 3 0 2 | 75.660 | 1.2560 | -0.251 | 0.0062 |
| 77.481 | 1.2336 | 2 2 0 | 77.675 | 1.2283 | -0.194 | 0.0053 |
| 79.687 | 1.2049 | 2 1 3 | 79.884 | 1.1998 | -0.197 | 0.0051 |
| 81.224 | 1.1860 | 3 1 0 | 81.173 | 1.1840 | 0.051 | 0.0020 |
| 83.656 | 1.1576 | 3 1 1 | 83.840 | 1.1530 | -0.184 | 0.0046 |

| Lattice parameter | Angle |
|---|---|
| a (Å) = 4.920 | α (°) = 90.00 |
| b (Å) = 4.920 | β (°) = 90.00 |
| c (Å) = 5.420 | γ (°) = 120.00 |

### ii. Biotite

| | | |
|---|---|---|
| Symmetry: Cubic | Space group = C12/m1 | COD ID: |
| Wavelength CuKα = 1.5406 nm | | Ref. cell volume = 489.14 Å³ |
| Wavelength CuKβ = 1.5444 nm | | Refined cell volume = 476.10 Å³ |

*************************************************************************

| OBSERVED | | | CALCULATED | | DIFFERENCE | |
|---|---|---|---|---|---|---|
| 2 Theta | d | h k l | 2 Theta | d | 2Theta | d |
| 9.009 | 9.8300 | 0 0 1 | 8.898 | 9.9300 | 0.111 | -0.1000 |
| 19.696 | 4.5137 | 0 1 2 | 19.982 | 4.4400 | -0.286 | 0.0737 |
| 23.389 | 3.8087 | 1 1 1 | 23.022 | 3.8600 | 0.367 | -0.0513 |
| 27.015 | 3.3052 | 0 0 3 | 26.994 | 3.3100 | 0.021 | -0.0048 |
| 35.078 | 2.5618 | -1 2 3 | 35.023 | 2.5600 | 0.055 | 0.0018 |

### iii. Pyrite

Symmetry: Cubic  Space group = Pa-3  COD ID: 1544891
Wavelength CuKα = 1.5406 nm  Ref. cell volume =159.040 Å³
Wavelength CuKβ = 1.5444 nm  Refined cell volume = 158.956 Å³
*****************************************************************

| OBSERVED | | | CALCULATED | | DIFFERENCE | |
|---|---|---|---|---|---|---|
| 2 Theta | d | h k l | 2 Theta | d | 2Theta | d |
| 28.453 | 3.1413 | 1 1 1 | 28.513 | 3.1280 | -0.060 | 0.0133 |
| 33.032 | 2.7156 | 2 0 0 | 33.084 | 2.7055 | -0.052 | 0.0101 |
| 40.761 | 2.2168 | 2 1 1 | 40.784 | 2.2107 | -0.023 | 0.0061 |
| 47.379 | 1.9215 | 2 2 0 | 47.411 | 1.9160 | -0.032 | 0.0055 |
| 56.261 | 1.6374 | 3 1 1 | 56.279 | 1.6333 | -0.018 | 0.0041 |
| 64.240 | 1.4519 | 3 2 1 | 64.283 | 1.4479 | -0.043 | 0.0040 |
| 71.727 | 1.3177 | 4 1 0 | 71.773 | 1.3141 | -0.046 | 0.0036 |
| 78.905 | 1.2149 | 4 2 0 | 78.962 | 1.2115 | -0.057 | 0.0034 |
| 88.280 | 1.1085 | 4 2 2 | 88.290 | 1.1060 | -0.010 | 0.0025 |

| Lattice parameter | Angle |
|---|---|
| a (Å) = 5.417 | α (°) =  90.00 |
| b (Å) = 5.417 | β (°) =  90.00 |
| c (Å) = 5.417 | γ (°) =  90.00 |

### iv. Titanium oxide

Symmetry: Monoclinic  Space group = A2/m  COD ID: 1100042
Wavelength CuKα = 1.5406 nm  Ref. cell volume =215 Å³
Wavelength CuKβ = 1.5444 nm  Refined cell volume = 216 Å³
*****************************************************************

| OBSERVED | | | CALCULATED | | DIFFERENCE | |
|---|---|---|---|---|---|---|
| 2 Theta | d | h k l | 2 Theta | d | 2Theta | d |
| 37.143 | 2.4230 | 0 3 1 | 37.217 | 2.4140 | -0.074 | 0.0090 |
| 43.610 | 2.0783 | -2 4 0 | 43.671 | 2.0710 | -0.061 | 0.0073 |
| 63.405 | 1.4691 | -2 4 2 | 63.493 | 1.4640 | -0.088 | 0.0051 |

| Lattice parameter | Angle |
|---|---|
| a (Å) = 5.855 | α (°) =  90.000 |
| b (Å) = 9.340 | β (°) =  90.000 |
| c (Å) = 4.142 | γ (°) = 107.530 |

### v. Magnetite

Symmetry: Orthorhombic  Space group = Bbmm  COD ID: 9002332
Wavelength CuKα = 1.5406 nm  Ref. cell volume =238.68 Å³
Wavelength CuKβ = 1.5444 nm  Refined cell volume = 235.26 Å³
*****************************************************************

| OBSERVED | | | CALCULATED | | DIFFERENCE | |
|---|---|---|---|---|---|---|
| 2 Theta | d | h k l | 2 Theta | d | 2Theta | d |
| 35.078 | 2.5617 | 1 1 1 | 35.284 | 2.5417 | -0.206 | 0.0200 |
| 38.764 | 2.3262 | 0 4 0 | 38.749 | 2.3220 | 0.015 | 0.0042 |
| 48.365 | 1.8845 | 3 2 1 | 48.251 | 1.8846 | 0.114 | -0.0001 |

| 2 Theta | d | h k l | 2 Theta | d | 2Theta | d |
|---|---|---|---|---|---|---|
| 52.601 | 1.7424 | 1 4 1 | 52.413 | 1.7443 | 0.188 | -0.0019 |
| 60.171 | 1.5400 | 3 4 1 | 59.949 | 1.5418 | 0.222 | -0.0018 |
| 61.024 | 1.5205 | 1 5 1 | 60.912 | 1.5197 | 0.112 | 0.0008 |
| 68.143 | 1.3780 | 0 0 2 | 67.984 | 1.3778 | 0.159 | 0.0002 |
| 78.971 | 1.2141 | 2 3 2 | 78.675 | 1.2152 | 0.296 | -0.0011 |
| 81.344 | 1.1845 | 4 1 2 | 81.791 | 1.1766 | -0.447 | 0.0079 |
| 87.303 | 1.1184 | 3 7 1 | 87.317 | 1.1158 | -0.014 | 0.0026 |

| Lattice parameter | Angle |
|---|---|
| a (Å) = 9.273 | α (°) = 90.00 |
| b (Å) = 9.239 | β (°) = 90.00 |
| c (Å) = 2.746 | γ (°) = 90.00 |

**Appendix B**
**Table (I):** Structural refinement parameters of all quartz samples and altered mineral after annealing
\*\*\*\*\*\*\*\*\*\*\*\*\*\*\*\*\*\*\*\*\*\*\*\*\*\*\*\*\*\*\*\*\*\*\*\*\*\*\*\*\*\*\*\*\*\*\*\*\*\*\*\*\*\*\*\*\*\*\*\*\*\*\*\*\*\*\*\*\*\*\*\*\*\*\*\*

i. **Quartz**

Symmetry: Hexagonal      Space group = P6222      COD ID: 1011200
Wavelength CuKα = 1.5406 nm            Ref. cell volume = 119 Å³
Wavelength CuKβ = 1.5444 nm            Refined cell volume = 119 Å³
\*\*\*\*\*\*\*\*\*\*\*\*\*\*\*\*\*\*\*\*\*\*\*\*\*\*\*\*\*\*\*\*\*\*\*\*\*\*\*\*\*\*\*\*\*\*\*\*\*\*\*\*\*\*\*\*\*\*\*\*\*\*\*\*\*\*\*\*\*\*\*\*\*\*\*\*

| OBSERVED | | | CALCULATED | | DIFFERENCE | |
|---|---|---|---|---|---|---|
| 2 Theta | d | h k l | 2 Theta | d | 2Theta | d |
| 20.639 | 4.3095 | 1 0 0 | 20.860 | 4.2550 | -0.072 | 0.0545 |
| 26.479 | 3.3709 | 1 0 1 | 26.640 | 3.3435 | -0.070 | 0.0274 |
| 36.197 | 2.4851 | 1 1 0 | 36.544 | 2.4569 | -0.106 | 0.0282 |
| 39.907 | 2.2622 | 1 0 2 | 39.465 | 2.2815 | -0.119 | -0.0193 |
| 42.029 | 2.1528 | 2 0 0 | 42.450 | 2.1277 | -0.113 | 0.0251 |
| 45.413 | 1.9999 | 2 0 1 | 45.793 | 1.9799 | -0.139 | 0.0200 |
| 49.816 | 1.8330 | 1 1 2 | 50.139 | 1.8180 | -0.148 | 0.0150 |
| 54.512 | 1.6857 | 2 0 2 | 54.875 | 1.6717 | -0.205 | 0.0140 |
| 59.434 | 1.5573 | 2 1 1 | 59.960 | 1.5415 | -0.168 | 0.0158 |
| 67.898 | 1.3823 | 2 0 3 | 67.744 | 1.3821 | 0.228 | 0.0002 |

| Lattice parameter | Angle |
|---|---|
| a (Å) = 5.015 | α (°) = 90.00 |
| b (Å) = 5.015 | β (°) = 90.00 |
| c (Å) = 5.4701 | γ (°) = 120.00 |

ii. **Hematite**

Symmetry: Cubic      Space group = R-3c      COD ID: 9009782
Wavelength CuKα = 1.5406 nm            Ref. cell volume = 159.22 Å³
Wavelength CuKβ = 1.5444 nm            Refined cell volume = 159.22 Å³
\*\*\*\*\*\*\*\*\*\*\*\*\*\*\*\*\*\*\*\*\*\*\*\*\*\*\*\*\*\*\*\*\*\*\*\*\*\*\*\*\*\*\*\*\*\*\*\*\*\*\*\*\*\*\*\*\*\*\*\*\*\*\*\*\*\*\*\*\*\*\*\*\*\*\*\*

| OBSERVED | | | CALCULATED | | DIFFERENCE | |
|---|---|---|---|---|---|---|
| 2 Theta | d | h k l | 2 Theta | d | 2Theta | d |
| 24.029 | 3.7087 | 0 1 2 | 24.144 | 3.6831 | -0.115 | 0.0256 |
| 33.103 | 2.7099 | 1 0 4 | 33.140 | 2.7010 | -0.037 | 0.0089 |
| 35.560 | 2.5281 | 1 1 0 | 35.636 | 2.5174 | -0.076 | 0.0107 |
| 49.323 | 1.8502 | 0 2 4 | 49.452 | 1.8416 | -0.129 | 0.0086 |
| 53.767 | 1.7070 | 1 1 6 | 54.042 | 1.6955 | -0.275 | 0.0115 |
| 62.398 | 1.4903 | 2 1 4 | 62.432 | 1.4863 | -0.034 | 0.0040 |
| 63.860 | 1.4597 | 3 0 0 | 64.011 | 1.4534 | -0.151 | 0.0063 |

Lattice parameter        Angle  
a (Å) = 5.420            α (°) = 55.28  
b (Å) = 5.420            β (°) = 55.28  
c (Å) = 5.420            γ (°) = 55.28